%% file: m84_ch08.tex
\documentclass{article}
\usepackage{emulateapj}
\usepackage{float,epsfig,psfig}
\usepackage{pstricks}
\input {apj_ref.tex}
\input {mydefs.tex}

\begin{document}

\submitted{ApJ Letters 2000, Submitted September 12, Accepted October 12}
\title{{Chandra Observation of M84, Radio Lobe Elliptical in Virgo
    cluster.}}  
\author{A.~Finoguenov$^{1,2}$ and C.~Jones$^2$}
\affil{
{$^1$ Max-Planck-Institut f\"ur extraterrestrische Physik,
             Giessenbachstra\ss e, 85748 Garching, Germany}\\
{$^2$ Smithsonian Astrophysical Observatory, 60 Garden st., MS 3, Cambridge,
  MA 02138, USA}}
\authoremail{alexis@head-cfa.harvard.edu}


\begin{abstract}
  
  We analyzed a deep {\it Chandra} observation of M84, a
  bright elliptical galaxy in the core of the Virgo cluster. We find that
  the spatial distribution of the soft X-ray emission is defined by the
  radio structure of the galaxy.  In particular we find two low density
  regions associated with the radio lobes and surrounded by higher density
  X-ray filaments. In addition to a central AGN and a population of galactic
  sources, we find a diffuse hard source filling the central 10 kpc
  region. Since the morphology of the hard source appears round and is
  different from that seen in the radio or in soft X-rays, we propose that
  it is hot gas heated by the central AGN. Finally, we find that the central
  elemental abundance in the X-ray gas is comparable to that measured
  optically.

\end{abstract}


\section{ Introduction}

M84 (NGC4374) is an E1 galaxy within the core of the Virgo cluster.  Radio
observations at 1.4 and 4.9 GHz show two lobes and a jet (Laing \& Bridle
1987).  In X-rays, the galaxy has a compact gas halo, rather low metallicity
and an overabundance of alpha-process elements (e.g. Finoguenov \& Jones
2000).

In this {\it Letter} we discuss primarily the comparison of the X-ray and
radio morphologies. We also determine the temperature and abundance
structures in the galaxy center, which allows a direct comparison with
optical abundance measurements. Finally, we resolve the nature of the hard
X-ray component, found by ASCA in many ellipticals (Matsumoto \etal 1997).

\section{Observation and data reduction} \label{sec:dr}

\begin{figure*}
\includegraphics[width=3.2in]{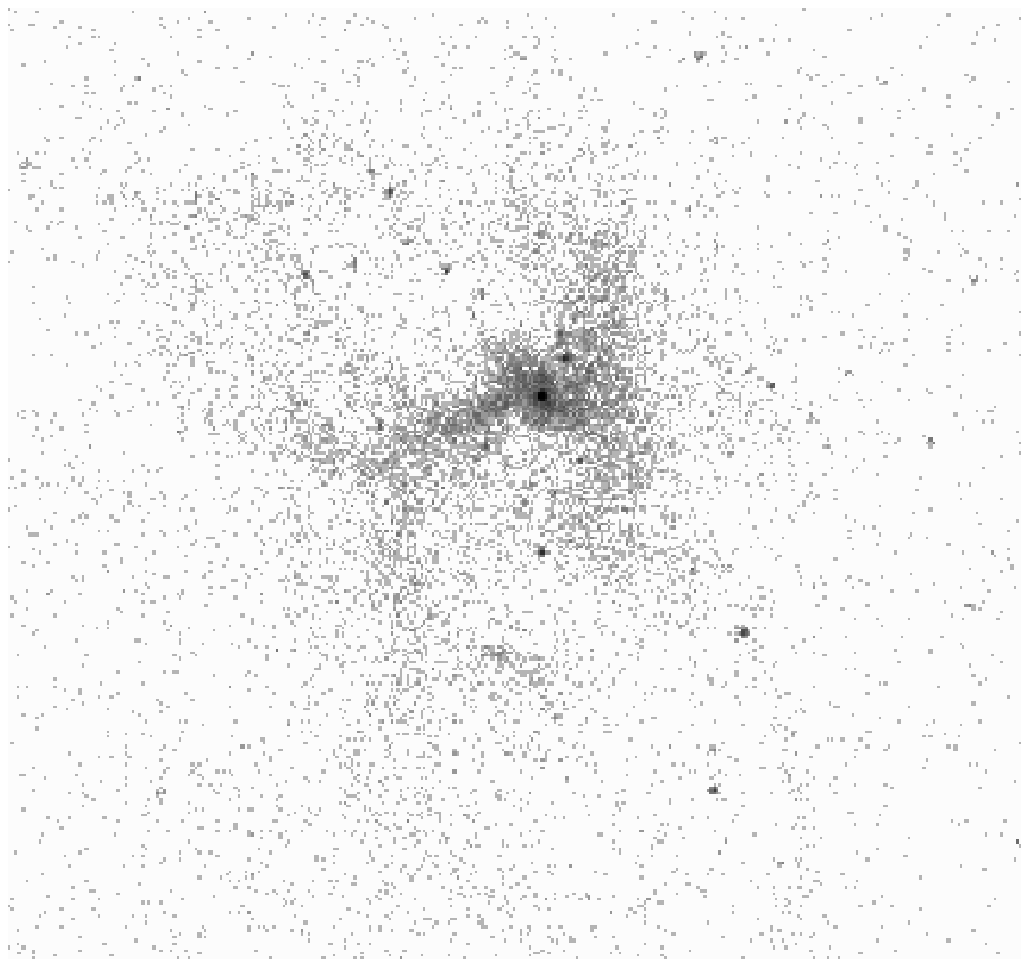}\hfill 
\includegraphics[width=3.2in]{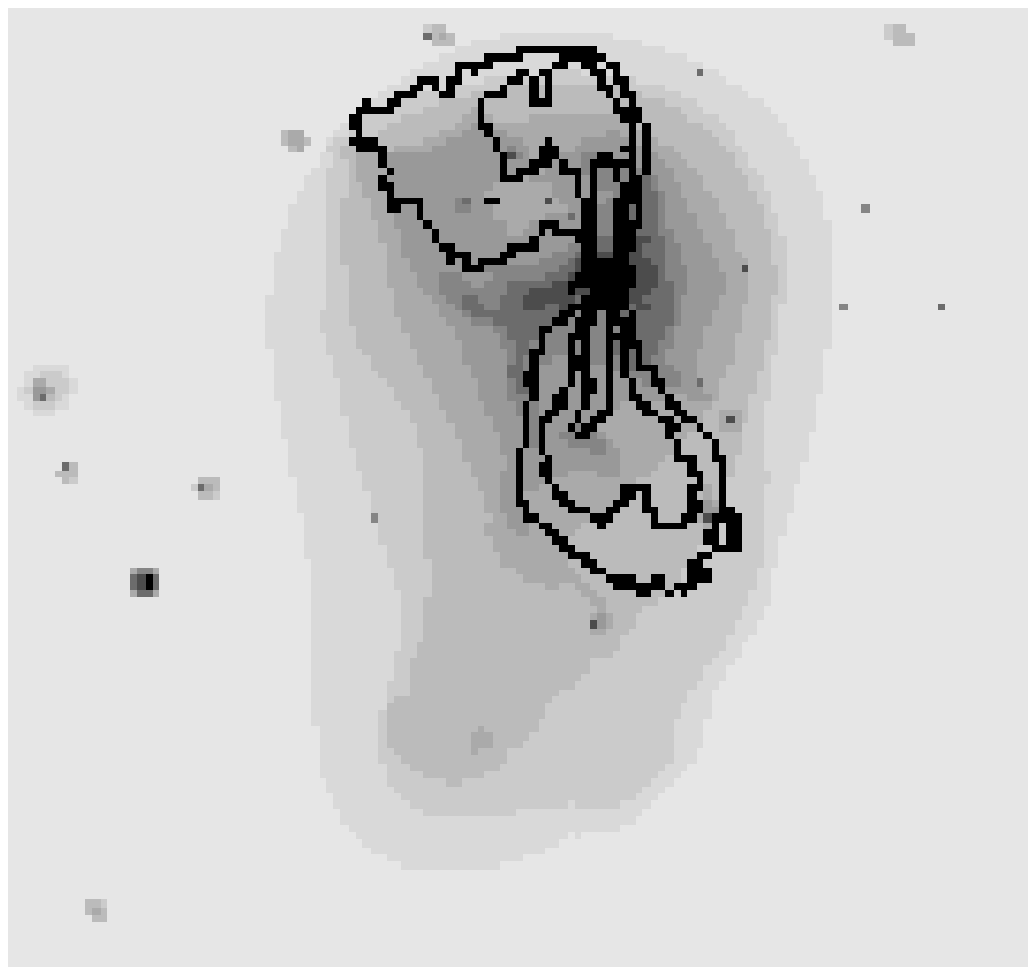}

\figcaption{Chandra M84 image in the 0.4--5.0 keV band {\it (left)} and
wavelet-decomposition overlaid with radio contours {\it (right)}. A 20\asec\
scale bar, shown on the images, corresponds to 1.6 kpc. Cross (on the left
image) denotes the center of radial profile extraction for southern
radio lobe.
\label{fig:img}
}
\pspicture(0.,9.)(0.,9.)
\psline(5.7,11.)(7.0,11.)
\rput(6.5,11.4){20\asec}
\psline(15.,11.)(15.6,11.)
\rput(15.3,11.4){20\asec}
\rput(4.9,11.9){x}
\endpspicture

\end{figure*}

A Chandra ACIS-S observation of M84 was made on 2000, May 19.
After removing time intervals with high X-ray background,
the useful exposure was 28.7 ksec. Gain corrections were made using
CIAO v1.1.1 software. Response matrices were generated for the focal plane
temperature of -120\deg C and position averaged, according to the source
location. Since the galaxy is embedded in diffuse X-ray
emission from the Virgo cluster, we calculated the background from a region
on the detector far from M84. We take the distance to M84 as 17 Mpc, so
1\asec\ corresponds to 82 pc. 

\section{Diffuse X-ray emission in M84}

Instead of appearing round, the bright extended X-ray emission from M84, shown
in Fig.\ref{fig:img}, has an ${\cal H}$-shape, characterized by a bar
extending east from the center and several almost parallel filaments
extending approximately perpendicular to this bar.

A clue towards understanding this peculiar X-ray morphology comes from a
comparison with the radio structure. Two radio lobes extend North and South
from the source's center. X-ray emitting gas surrounds the radio lobes. As
discussed in Sec.\ref{sec:kt-ab}, the temperature of the X-ray filaments is
0.6--0.7 keV, similar to that at the galaxy center and slightly colder than
the more extended X-ray emission, seen on scales up to 25 kpc.  Apparently,
expansion of the radio lobes has caused the hot gas to have the ${\cal
H}$-shaped structure. However, as will be shown in Sec.\ref{sec:kt-ab}, the
temperature of the filaments surrounding the radio lobes is not very
different from the temperature found for the central and outer region of the
source. Thus, we conclude that the lobes are not overpressured and therefore
are not capable of driving a strong shock into the ambient material. This
interpretation is similar to that for Hydra-A (McNamara \etal 2000).

We can use the density ($n_e$) of the hot gas surrounding the radio lobes,
determined from X-ray observations combined with radio measurements of the
Faraday rotation ($R_m$) to directly determine the strength of the magnetic
field ($B_{\|}$) (\eg\ Laing \& Bridle 1987).

\[ R_m \propto \int n_e B_{\|} dl \]

To determine the magnetic field strength we extracted radial profile of the
X-ray surface brightness around the geometrical center of the southern radio
lobe to model the 3-d density profile (see Fig.\ref{fig:pro}). We model this
with a $\beta$-model, modified by setting the density to zero inside some
radius, so the emission detected inside this radius is only due to the
projection effects.  Using the results of this modeling, we estimate the
integrated electron density to be $\sim 0.04$ cm$^{-3}$ kpc, which, for a
Faraday rotation of 25 rad m$^{-2}$, requires the line-of-sight magnetic
field strength to be 0.8 $\mu G$, well below the equipartition value of 20
$\mu G$ (Laing \& Bridle 1987). Thus, the magnetic field associated with the
X-ray plasma appears much weaker then the field in the radio lobe. This
method, however, has certain limitations, since on one hand, it sets an
upper limit on the magnetic field, due to the possible presence of plasma
between the source and the observer, and on the other hand, there is an open
question of which scale size of the magnetic field reversals is most
responsible for the observed RM and whether this scale is characteristic of
the strength of the magnetic field.

\includegraphics[width=2.8in]{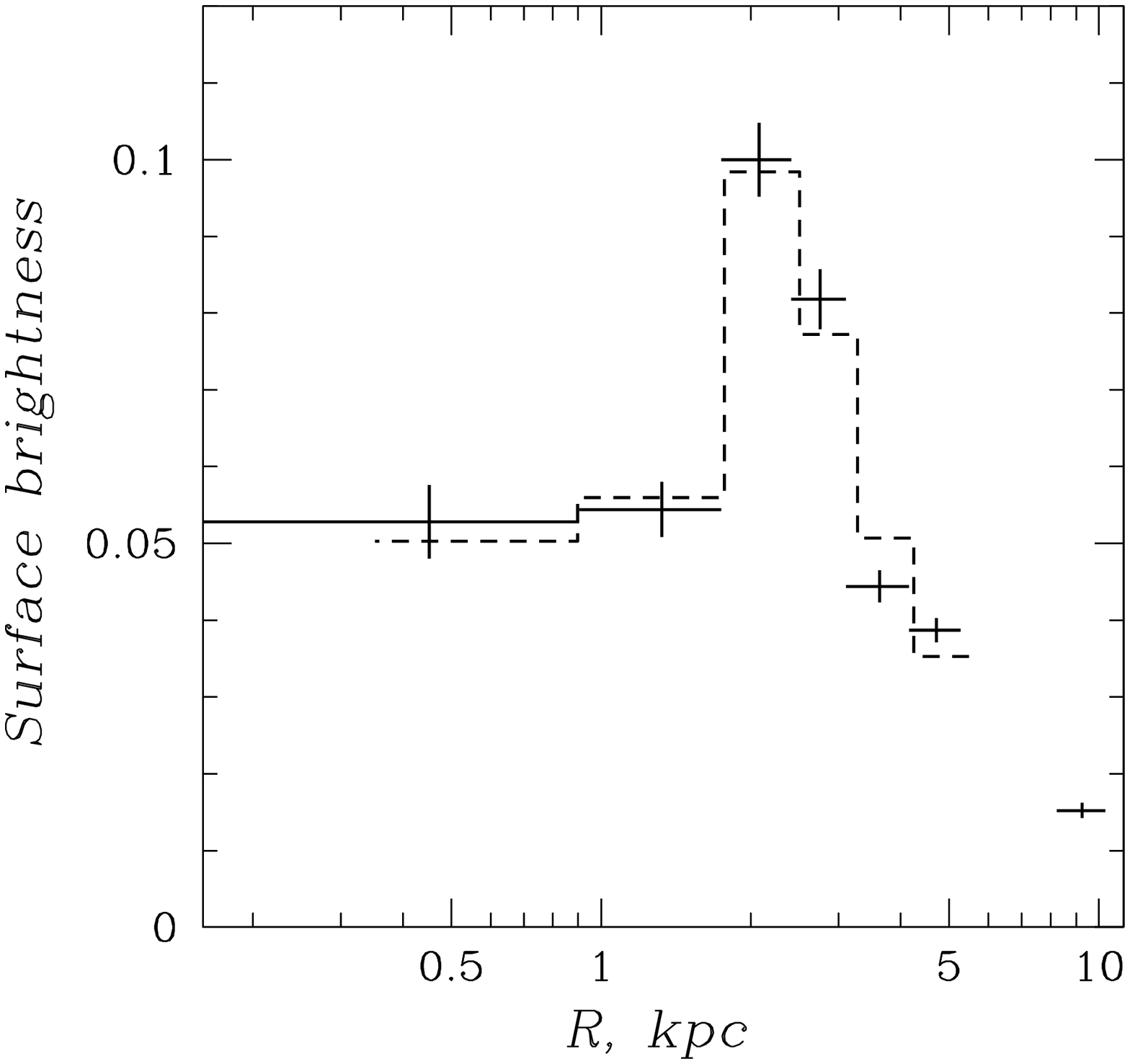} 

\figcaption{Radial X-ray surface brightness profile (in counts per pixel,
pixel size is $0.5^{\prime\prime}\times0.5^{\prime\prime}$) around the
southern radio lobe. A point at 10 kpc indicates the background level. A
dashed histogram shows the best fit for a spherical shell of gas.
\label{fig:pro}
}

The radio and X-ray morphologies suggest that M84 is moving
in the core of the Virgo cluster.  In particular 
the northern edge of the north radio lobe appears compressed
and, as shown on the right of Fig.\ref{fig:img}, although the
faint X-ray emission has a more regular shape, it is extended towards the
south.  Both the
X-ray extension and radio structure suggest motion through the Virgo cluster
towards the north-north-west in the observed plane.
We also note that the
radio jet, seen in the southern lobe, points to a cloudlet of X-ray emission
3.3 kpc (40\asec) from the center, which could be compressed plasma
swept up by the jet.  

\section{Temperature and Heavy Element Abundances}\label{sec:kt-ab}

Spectra for different regions of M84 are presented in Fig.\ref{fig:spe}.  In
our spectral modeling of thermal emission from M84, we use the MEKAL model
(Mewe \etal 1985, Mewe and Kaastra 1995, Liedahl \etal 1995). Unless noted
otherwise, we found the Galactic absorption column of $2\times10^{20}$
cm$^{-2}$ (Stark \etal 1992). Errors are given on the 68\% confidence
level. We determined abundances for O, Mg, Si and Fe. In deriving the
abundances, we use the solar abundance table from Anders \& Grevesse (1989),
where elemental number abundances for O, Mg, Si, Fe relative to hydrogen are
(85.1, 3.8, 3.55, 4.68)$\times10^{-5}$.
We chose the energy interval 0.6--7 keV for spectral analysis. We
masked out the point sources when extracting the spectra of the
extended emission and in spectral fitting allowed for the presence of the
hard X-ray emission. We detect little variation in the temperature ($0.60\pm0.05$ keV) for the
different parts of the ${\cal H}$-shaped emission with the east part 
slightly hotter and the south-east filament  hotter by 0.1
keV. In the diffuse emission, if we fit both hard and soft components, the
temperature of the soft component rises from 0.6 keV at the center to 0.8
keV at 5\amin\ radii. 

We specify four zones, where we determine the abundances of heavy elements:
the inner 6\asec\ (0.5 kpc), the eastern and western parts of ${\cal
H}$-shaped diffuse emission and the faint X-ray halo, extending from 2\amin\
to 5\amin. In Fig.\ref{fig:ab} we plot the results for the ${\cal H}$-shaped
emission from 6\asec\ to 1\amin\ (east) and from 1\amin\ to 2\amin\
(west). The abundance determination is sensitive to the background
subtraction and is less reliable for the faint X-ray halo. Among elements,
Si abundance is most strongly affected, while the Mg abundance is the most
robust.

The ${\cal H}$-shaped region contributes most of the diffuse X-ray
emission and thus provides the tightest constraints on measurements of
the elemental abundance. The Fe abundance is highest in the ${\cal
H}$-shaped diffuse emission and highest in the eastern part. The Fe
abundance in ${\cal H}$-shaped region is well in excess of -0.35 dex.,
found in optical studies (Kobayashi \& Arimoto 1999), allowing for
additional gas enrichment from SN Ia.  The Mg/Fe ratio shows a trend
toward an overabundance of alpha-process elements at low
metallicities.  In the faint X-ray halo, low-metallicity inhibition of
the SN Ia rate is needed to explain the low Fe abundance. Low Fe abundance
in the center could be an artifact, due to the possible presence of
additional emission components, \eg\ X-ray scattering of AGN
emission on the galactic dust.
Our findings demonstrate that the previously reported disagreement between
X-ray and optical abundance estimates was due to their different spatial
scales of measurements.

\includegraphics[width=3.2in]{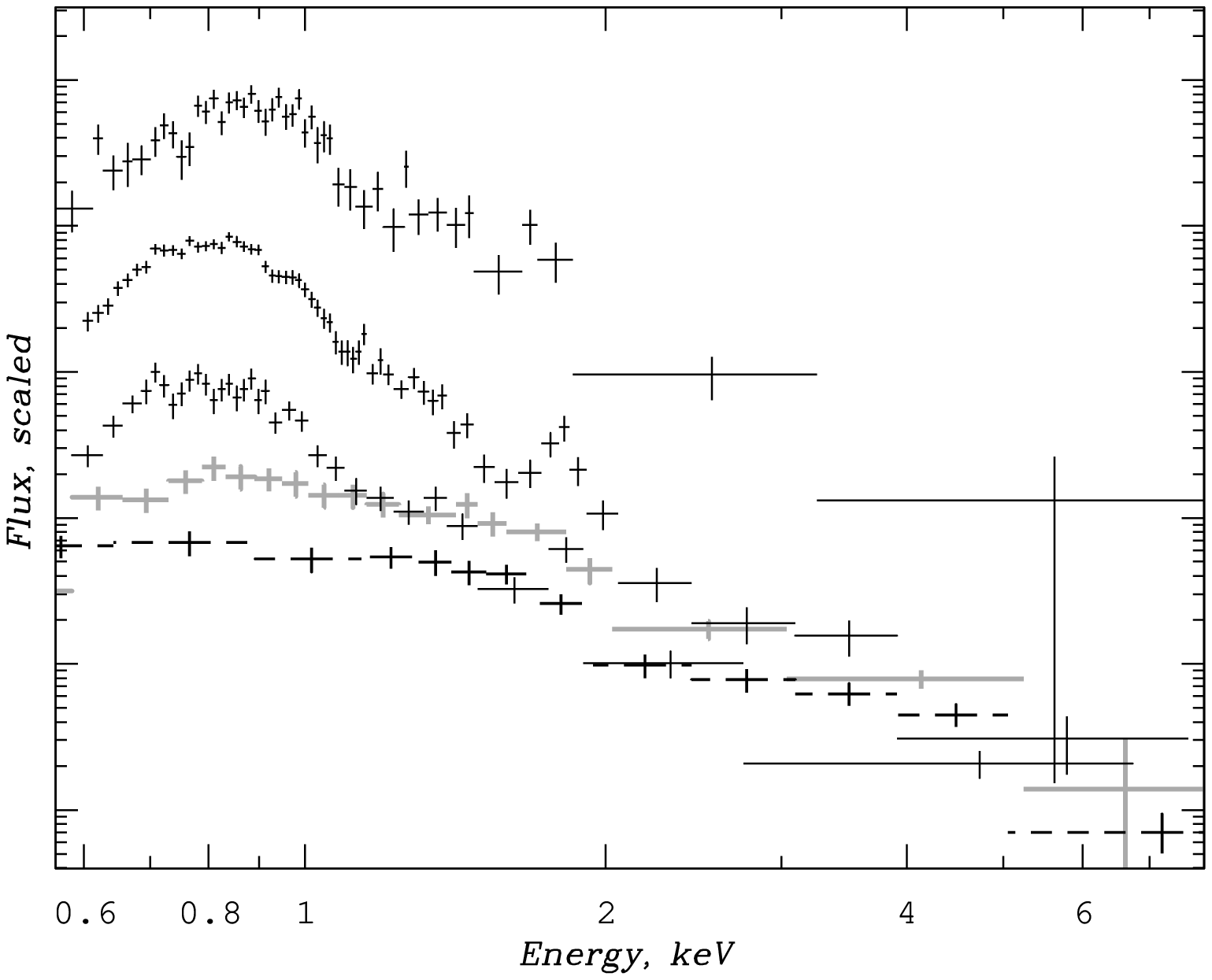} 

\figcaption{Spectral characteristics for regions of M84, representing from
  bottom to top, combined point sources (dashed line), the central AGN (in
  grey), central diffuse emission, ${\cal H}$-shaped region, outer faint
  X-ray emission.
\label{fig:spe}
}

\includegraphics[width=3.2in]{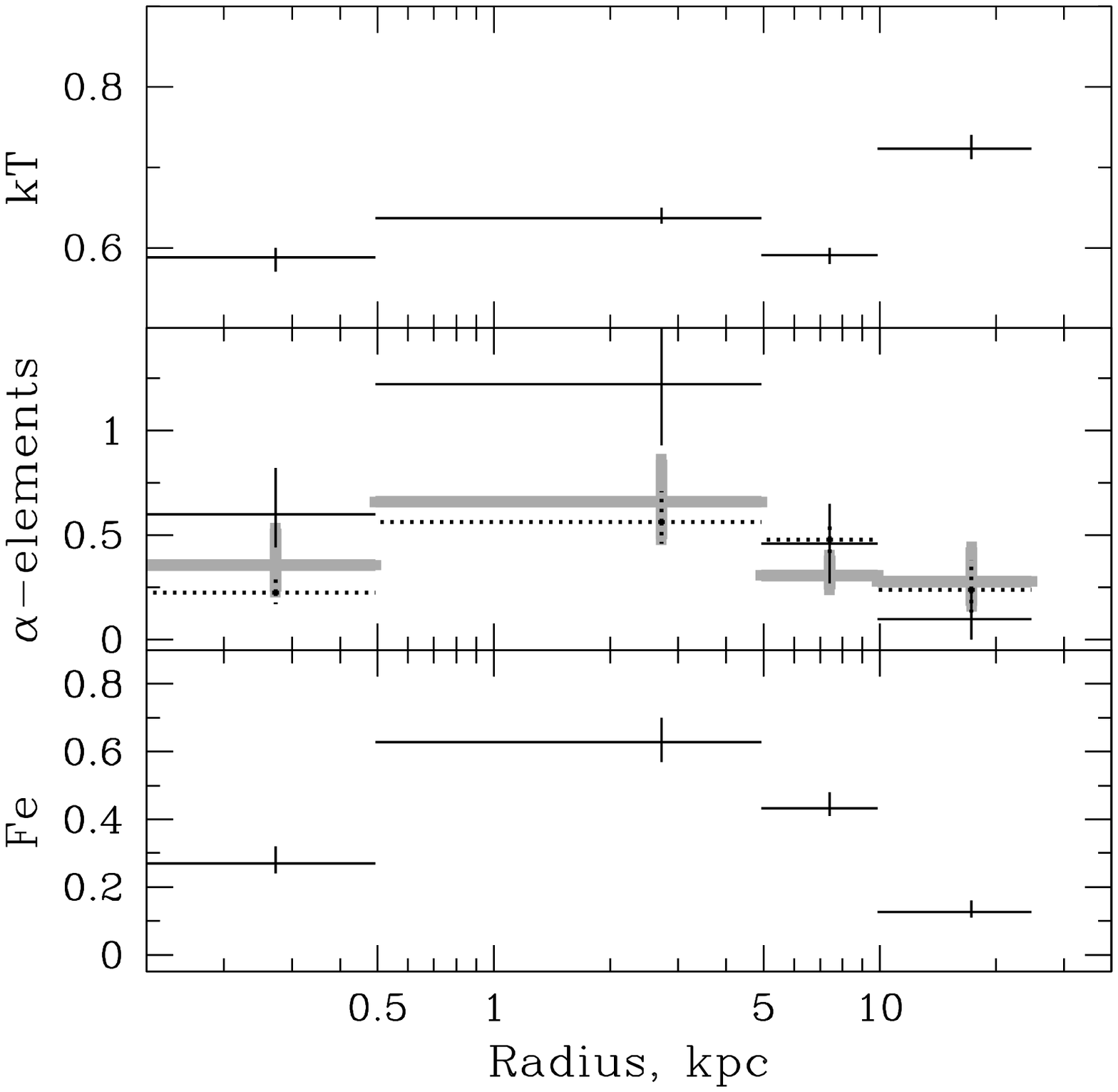} 

\figcaption{Radial distribution of temperature and heavy element abundances
  in M84. Among $\alpha$-process elements we plot Mg (in grey), Si (in solid
  black) and O (in dotted black).
\label{fig:ab}
}

\section{Origin of the Hard Component}

A hard X-ray component in elliptical galaxies has been found in many ASCA
observations (\eg Matsumoto \etal 1997) with spectral characteristics
corresponding to a 6.5 keV bremsstrahlung temperature. 
Three origins have been proposed to explain the hard component. First --
LMXB (Matsumoto \etal 1997), second -- AGN (\eg\ Finoguenov \& Jones 2000)
and third -- the presence of relativistic particles, contained by a magnetic
field, introduced by Fukazawa \etal (2000) to explain the hard component
seen in groups of galaxies. Chandra's subarcsecond resolution allows us to
distinguish among these three possibilities, since the first is an ensemble
of discrete sources, the second is a point source in the galaxy center and
the third is a diffuse source.

{
\includegraphics[width=2.8in]{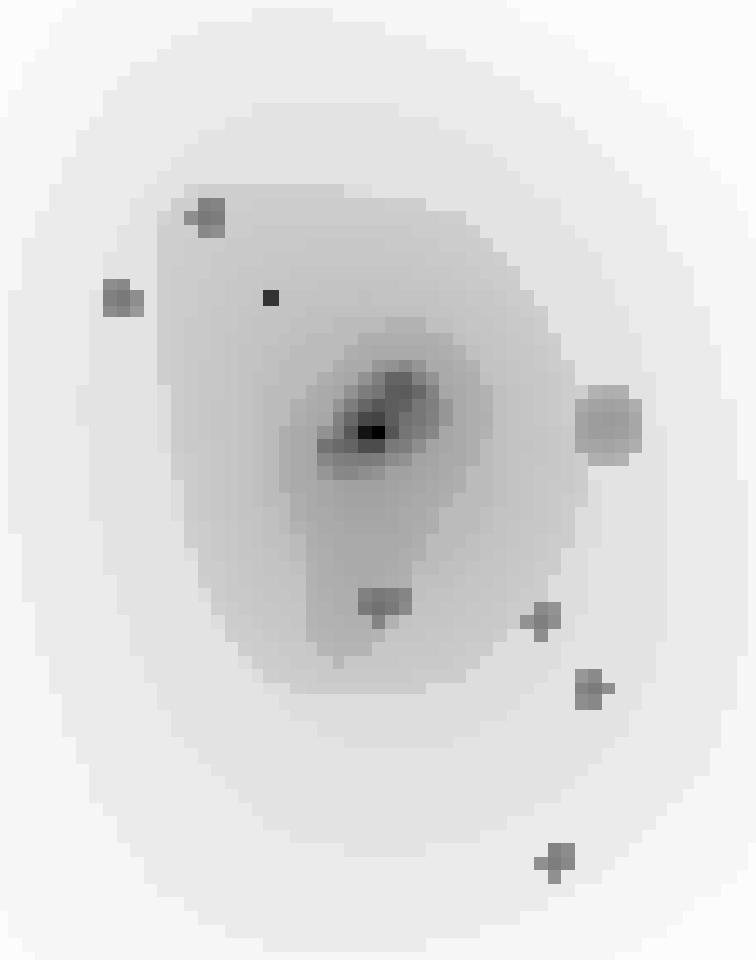} 
\figcaption{ Wavelet-decomposition of M84 image at 2.5--7.0 keV energy band.
  In contrast to the soft image, the hard emission does not reveal the
  ${\cal H}$-shape, instead is spherically symmetric, extending to 3.5 kpc
  radius from the center. At the source center, emission exhibits a
  distinct bar pointed to the north-west. A 20\asec\
scale bar, shown on the image, corresponds to 1.6 kpc.
\label{fig:hard}
}
\pspicture(0.,9.)(0.,9.)
\psline(5.7,12.)(7.0,12.)
\rput(6.5,12.4){20\asec}
\endpspicture
}

Fig.\ref{fig:hard} shows the 2.5--7 keV emission. In this hard band, in 28.7
ksec, the diffuse emission provides 240 counts, the central point source 100
counts, with 167 counts from detected point sources.

The central X-ray point source is coincident with the position of the radio
nucleus. Its spectrum is an absorbed ($N_H=2.7\; (2.4-3.0)\times 10^{21}$
cm$^{-2}$) power law ($\Gamma=2.3\; (2.2-2.4)$) and its luminosity is
$4\times10^{39}$ ergs/sec in the 0.5--10 keV band.  The integrated spectrum
of the combined emission from point sources is characterized by a
bremsstrahlung temperature of 13 (9--22) keV or a power law index 1.41
(1.34--1.48). Their total luminosity in the 0.5--10. keV band is
$8\times10^{39}$ ergs/s, yielding an $L_X/L_B$ ratio similar to M31
(Makishima \etal 1989).

To estimate the contribution of point sources below our detection limit, we
study the point source $log(N)-log(S)$ distribution in the 0.4--5.0 keV
band, using the central $20\times20$ kpc$^2$ region. A fit to the source
counts distribution, excluding the first and the last bin of the histogram
shown in Fig.\ref{fig:src-hist}, gives $N=36\times S^{-0.43}$. Assuming no
break at lower source fluxes, we estimate an additional 1/3 of the point
source flux in undetected sources. Thus, point sources cannot explain the
hard diffuse emission in M84.  We note that at our point source sensitivity
limit (10 cnts detected in the 0.4--10. keV band correspond to
$8\times10^{-16}$ erg cm$^{-2}$ s$^{-1}$ in the 0.5--2 keV band for the
average slope found for bright background sources of $\Gamma=1.7$), using
the log N--log S determination by Giacconi \etal (2000), 3 of the 50 sources
we detect should be background; one identified quasar is omitted from the
source list.

The diffuse hard component extends to $\sim10$ kpc from the galaxy
center. For spectral estimates, however, we take the central bright 3.1 kpc
radius, where the uncertainty in the background subtraction is less
important. In the spectral analysis, we also fit the soft diffuse
component. The hard X-ray component can be satisfactory described by a
thermal plasma of 3.9 (2--11) keV giving a luminosity of $1.2\times10^{40}$
ergs/sec in the 0.5--10 keV band. A power law fit to the hard diffuse
component gives $\Gamma$ of $2.0\; (1.6-2.2)$.

Fukazawa \etal (2000) suggested that hard diffuse emission is due to the
Inverse Compton Scattering of CMB photons on relativistic electrons.
Alternatively, they suggested a non-thermal bremsstrahlung between thermal
gas and high energy particles. Both these scenarios have difficulties
explaining the observed shape of the diffuse emission in M84, since the hard
X-ray emission should correlate well with the radio emission in the first
scenario (\eg\ Sarazin 1988), and with the soft X-ray emission in the
second. Both radio and soft X-ray morphologies are well defined and do not
correlate with the hard X-ray component. However, as we noted earlier, 
the magnetic field strength outside the radio lobes is 25
times lower, which implies a reduction in the synchrotron emission by almost
three orders of magnitude.  Therefore, to explain a mismatch
between the radio and hard X-ray diffuse emission, energetic electrons with
estimated $\gamma=10^{3-4}$ must be distributed similar to the X-ray
emission. Thus, differences in X-ray and radio morphology could be due to the
change in the magnetic field strength. The electrons, required for this
scenario would have life-times of $\sim10^9$ yr, so their supply is not a
problem (Fukazawa \etal 2000).

As an alternative to an Inverse Compton origin, we suggest that the hard
diffuse X-ray emission is due to AGN-inflated gas in the center of the
galaxy. In order for this suggestion to hold, several requirements must be
met. First, in order for the hot gas to be convectively stable (\eg\ Sarazin
1988), the entropy of the hot gas should not exceed the entropy of the faint
diffuse halo. The latter has an entropy of $90\pm20$ keV cm$^{2}$ (ignoring
constants and logarithms). For the measured density of the hard component,
this requires that the temperature be less then 4.6 keV, consistent with
that measured 3.9 (2--11) keV. Hotter temperatures are allowed closer than
0.8 kpc from the center, where the gas density is higher. The second
requirement is that the gas be in pressure equilibrium in the gravitational
field of M84. In fact, we would expect there to be a hole at the center of
the soft emission, which in M84 is masked by the complex structure in the
central region. The spatial distribution of the hard component is
characterized by a $\beta$-model with $\beta= 0.43\pm0.04$, $r_c=0.97\pm0.12$
kpc. Assuming an isothermal temperature of 3 keV, the total gravitational
mass within 3.1 kpc is $4.0\times10^{11}$\msun.  Using the cooler gas
component ($\beta=1.40\pm0.03$, $r_c=5.28\pm0.08$ kpc) at larger radii, we
find $9.2\times10^{11}$\msun\ within 10 kpc. For a total blue luminosity of
$5\times10^{10}$\lsun\ and effective radius of 5 kpc (Kobayashi \& Arimoto
1999), both mass estimates correspond to a $M/L_B$ ratio of 20. This is
similar to Brighenti \& Mathews (1998) result of constant $M/L_B$ in
ellipticals over a large range of radii. This is consistent with the
assumption that the gas is in pressure equilibrium. Finally, the total
energy of the hot component is $\sim10^{56}$ ergs, which requires a heat
input of $\sim10^{41}$ ergs/s for 0.1 Gyr. Future X-ray observations can
test this suggestion through a more precise temperature determination and
the detection of Fe K-shell lines.

\includegraphics[width=2.4in]{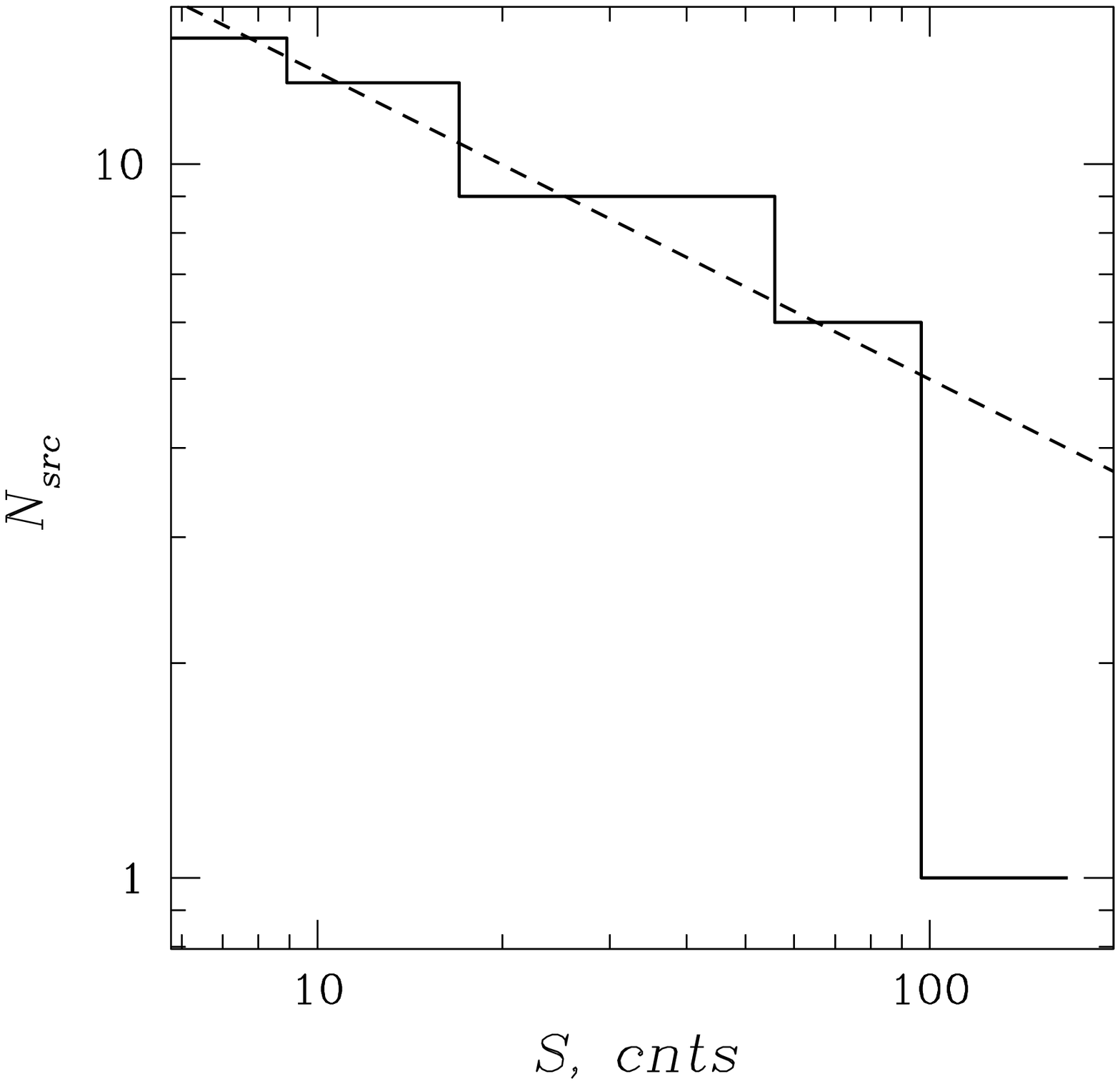}
\figcaption{$log(N)-log(S)$ of the point source counts in 0.4--5.0 keV
energy band, using the central $20\times20$ kpc$^2$ region (solid
histogram). Dashed line shows the best fit. Given the duration of the
observation and a source spectral shape, 1 cnt on this plot corresponds to
$10^{37}$ ergs/sec luminosity in 0.4--10. keV.
\label{fig:src-hist}
}

In the proposed scenario, the cooler X-ray emitting filaments surrounding
the radio-lobes will tend to fall toward the center, due to their lower
entropy, but are repelled by the pressure in the radio-lobes, which
apparently has little impact on the diffuse hot X-ray gas.  Thus, for the
cool component of the X-ray emission, the radio source works as a fountain,
lifting up the infalling gas.

\section{Conclusions}

The high resolution Chandra image shows the remarkable interaction of the
radio lobes and the diffuse soft X-ray emission. The expansion of the radio
lobes appears to have created cavities in the soft emission that are
surrounded by higher density shells. Based on the similarities in the
temperature and abundance of the filaments around the radio lobes and the
gas in the central region, we suggest that this gas has a common origin and
circulates throughout the central region of the galaxy. Chandra's high
resolution also has led to the detection of a population of luminous
($>10^{38}$ ergs/sec) galactic X-ray sources and the discovery of a hard,
diffuse source in the central 10 kpc. We suggest that the AGN at the center
of M84 heats the gas in this region. Future observations of M84 and other
galaxies can test this prediction. Finally, Chandra's resolution allowed
abundance measurements to be made on a spatial scale similar to those done
optically, resolving the issue of low reported abundances for the X-ray gas,
compared to optical studies.

\vspace*{0.2cm}

The authors are thankful to Alexey Vikhlinin and Maxim Markevitch for their
help in Chandra data reduction, and to Dan Harris, Yasushi Fukazawa, Brian
McNamara and the referee for useful comments on the manuscript. This work
was supported by NASA grants GO0-1045X and AG5-3064 and the Smithsonian
Institution.

\end{document}

%% file: apj_ref.tex
   \def\apj#1#2#3#4{\par #4 19#3, {ApJ,\/} {#1}, #2 }

   \def\apjs#1#2#3#4{\par #4 19#3, {ApJ (Supplement Series),\/} { #1}, #2 }

   \def\aas#1#2#3#4{\par #4 19#3, {A\&A (Supplement Series),\/} { #1}, #2 }
   
   \def\mnras#1#2#3#4{\par #4 19#3, {MNRAS,\/} { #1}, #2 }

   \def\apjl#1#2#3#4{\par #4 19#3, {ApJ (Letters),\/} { #1}, #2 }

   \def\BIB {\par}

%% file: mydefs.tex
\def\hea4{{\it HEAO~A4}}
\def\heaoa2{{\it HEAO~A2}}
\def\heao1{{\it HEAO~1}}

\def\amin{$^\prime$}
\def\asec{$^{\prime\prime}$}

\def\eg{{\it e.g.}~}

\def\deg{$^{\circ}$}

\def\h0{$H_{\rm o}=50$~km~s$^{-1}$~Mpc$^{-1}$}
\def\q0{$q_{\rm o}$}

\def\msun     {$M_{\odot}$}
\def\lsun     {$L_{\odot}$}

\def\etal    {{ et~al.}~}

\def\cms3  {~{cm$^{-3}$}}